\documentclass[aps,prl,preprint,groupedaddress,showpacs,amsmath,amssymb,onecolumn,floatfix]{revtex4-1}
\usepackage{graphicx}
\usepackage{amsmath, amsthm, amssymb}
\usepackage{latexsym}
\usepackage{amssymb}
\usepackage{bm}
\usepackage{dcolumn}
\usepackage{epstopdf}
\usepackage{color}
\usepackage{verbatim}

\begin{document}

\title{Piezoacoustics for precision control of electrons floating on helium}

\author{H. Byeon$^1$}
\email[]{byeonhee@msu.edu}
\author{K. Nasyedkin$^1$}
\author{J.R. Lane$^1$}
\author{N.R.~Beysengulov$^1$}
\author{L. Zhang$^1$}
\author{R. Loloee$^1$}
\author{J. Pollanen$^1$}
\email[]{pollanen@msu.edu}
\affiliation{$^1$Department of Physics and Astronomy, Michigan State University, East Lansing, MI 48824, USA}

\date{\today}

\begin{abstract}
Piezoelectric surface acoustic waves (SAWs) are powerful for investigating and controlling elementary and collective excitations in condensed matter. In semiconductor two-dimensional electron systems SAWs have been used to reveal the spatial and temporal structure of electronic states, produce quantized charge pumping, and transfer quantum information. In contrast to semiconductors, electrons trapped above the surface of superfluid helium form an ultra-high mobility, two-dimensional electron system home to strongly-interacting Coulomb liquid and solid states, which exhibit non-trivial spatial structure and temporal dynamics prime for SAW-based experiments. Here we report on the coupling of electrons on helium to an evanescent piezoelectric SAW. We demonstrate precision acoustoelectric transport of as little as $\sim0.01\%$ of the electrons, opening the door to future quantized charge pumping experiments. We also show SAWs are a route to investigating the high-frequency dynamical response, and relaxational processes, of collective excitations of the electronic liquid and solid phases of electrons on helium.
\end{abstract}

\maketitle
\section{Introduction}
The surface of superfluid helium at low temperatures is a fantastically pristine substrate without the defects that are unavoidable in almost all other material systems. Electrons placed near this superfluid substrate are attracted to it and float $\sim$~10 nm above the surface, forming a unique two-dimensional electron system (2DES) with the highest electron mobility in condensed matter~\cite{Shirahama1995}. Electrons on helium are quantum non-degenerate, but not classical, with quantum effects influencing the many-electron transport in high magnetic field~\cite{Dykman1997} as well as the single electron degrees of freedom. 

Strong repulsive forces between electrons in this system conspire to produce electronic fluid~\cite{Lea97} or crystallized electronic solid states~\cite{Gri79} exhibiting exotic high-frequency dynamical response~\cite{Konstantinov2010,Chepelianskii2015,Abdurakhimov2016,Yunusova2019,Kawakami2019a,Zadorozhko2021}, which is typically investigated via free-space coupling between the electrons and radio-frequency or microwave fields. Surface acoustic wave (SAW) based techniques are a promising complementary probe for investigating these high-frequency collective phenomena. The wavelength of a SAW can be made comparable to the inter-electron spacing in this system, opening up SAW-based scattering measurements of the Wigner solid state of electrons on helium~\cite{Williams1998} or engineering acoustic lattices~\cite{Schuetz2017} for investigating electron crystallization in the presence of a commensurate trapping potential~\cite{Moskovtsev2020}. 

The superfluid substrate is also predicted to facilitate slow electron decoherence, which has attracted interest to electrons on helium as a candidate for quantum information processing~\cite{Platzman1999,Lyon2006,Sch10,Schuster2016,Kawakami2019a,koo19}. In this context, recent circuit quantum electrodynamic (cQED) experiments have studied the motion of a single electron on helium coupled to a microwave resonator~\cite{koo19}. Additionally, advances in micro- and nano-fabrication techniques allow for precision control of electrons on helium in confined geometries and mesoscopic devices. Single-electron detection has been experimentally achieved~\cite{Papageorgiou2005} along with stable, high-fidelity, electron transfer along gated CCD arrays~\cite{Bradbury2011}. Point-contact devices~\cite{Rees2011}, mesoscopic field effect transistors (FETs)~\cite{Ashari2012a,Shaban2016}, and quasi-1D microchannels~\cite{Rees2016} have demonstrated a rich variety of precision transport capabilities in this 2DES on helium. To ultimately realize coherent control of single electrons will likely require devices that bring together multiple of these experimental advances. Given their broad success when coupled with semiconductor systems~\cite{Paa92,Wil93,Kuk11,Bertrand2016,Pol16,Friess2017,Takada2019,Kataoka2009} and quantum circuits~\cite{Delsing2019}, it is natural to ask if SAW techniques could be employed to manipulate electrons on helium and added to the toolkit for precision control of this unique low-dimensional electron system.

Surface acoustic waves on piezoelectric crystals resemble microscopic earthquake-like excitations accompanied by electric fields localized to a region approximately one wavelength above and below the crystal surface. These co-propagating electric fields interact with mobile charges located in close proximity to the piezoelectric surface wave. Since these waves propagate with a speed of several thousand meters per second standard optical and electron beam fabrication techniques are routinely used to create SAW devices with frequencies ranging from 100's of MHz to several GHz. Of particular interest is the transfer of momentum from a traveling SAW to nearby charges to produce controlled charge pumping~\cite{Shil95,Shilton1996,Tal97,Lane2018}. In this case the SAW induces a traveling oscillating piezoelectric field that can trap individual electrons, which then propagate in the local minima of the traveling SAW potential (see Supplemental Information Section 2 for a detailed description of the microscopic interaction between the SAW and the system of electrons on helium). Theoretically the interaction of SAWs with electrons on helium was proposed and investigated in Ref.~\cite{Wilen1986}, but to our knowledge no acoustoelectric experiments have been demonstrated with electrons on helium until now.

In this manuscript we report on experiments coupling electrons on superfluid helium-4 to a piezoacoustic SAW-field. The results demonstrate a key step, namely that SAW methods are compatible with electrons on helium and can be used to produce controlled acoustoelectric charge transport. This work paves the way for a class of SAW-based experiments ranging from fundamental studies of the structure of the electron system and its collective excitations to possible hybrid systems coupling small numbers of electrons to quantum acoustic devices~\cite{Gus14,Are16}.

\section{Results \& Discussion}

\begin{figure}[hbtp]
\centerline{\includegraphics[height=0.4\textheight]{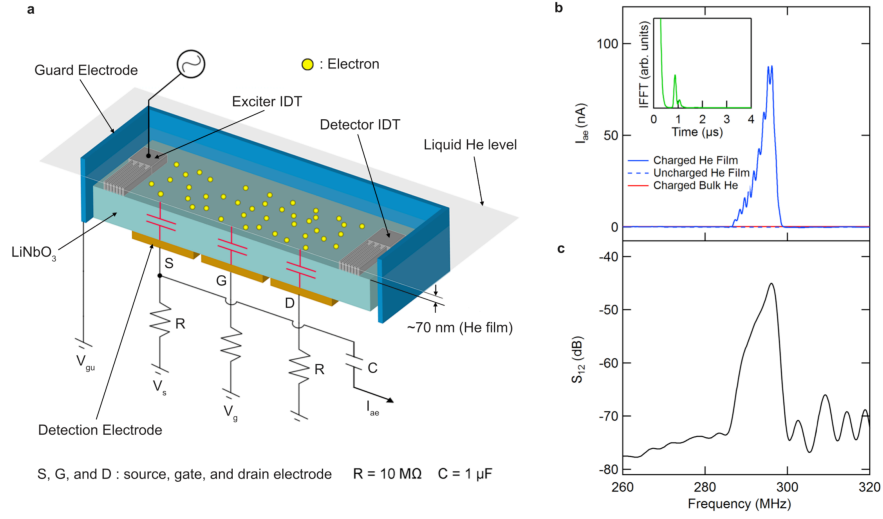}}
\caption{{\bf Schematic of the experimental setup and demonstration of continuous wave acoustoelectric transport of electrons on helium.} \textbf{a)} Cross-section view of the device for measuring SAW-driven transport of electrons on helium. Two opposing interdigitated transducers (IDTs) are used to excite and receive SAWs. A saturated superfluid \textsuperscript{4}He film is formed on the surface of the LiNbO\textsubscript{3} piezo-substrate at $T = 1.55$ K. Thermionically emitted electrons are trapped above the surface of the superfluid film by applying positive bias voltages to three underlying electrodes arranged in a field-effect transistor (FET) configuration with a source (s), gate (g) and drain (d)~\cite{Nas18}. Lateral confinement of the electron layer is achieved with a negative bias to the guard electrode positioned on the outside of the LiNbO\textsubscript{3} substrate. \textbf{b)} Measured acoustoelectric current $I_{\text{ae}}$ of electrons on helium driven by a piezo-SAW as a function of frequency. For these measurements the FET electrode voltages were $V_{\text{s}}$ = $V_{\text{g}}$ = $V_{\text{d}}$ = 40~V, corresponding to an electron density of $n \cong 0.8 \times 10^9$~cm\textsuperscript{-2}, and the guard was biased with $-3.2$ V. Inset: Inverse Fourier transform of the acoustoelectric current signal which reveals a peak at $t \approx$ 0.9 $\mu$s. This time scale corresponds to a SAW propagation distance of 3.2 mm, roughly the same as twice the distance between the launching IDT center and the near-edge of the LiNbO\textsubscript{3} substrate, which implicates SAW reflections as responsible for the superimposed oscillations present on the acoustoelectric current peak. \textbf{c)} Frequency dependence of the transmission coefficient ($S_{12}$) of the SAW device demonstrating an expected resonance at 296~MHz.}
\label{fig1}
\end{figure}

Fig.~\ref{fig1}(a) shows a schematic of the device for producing and measuring SAW-driven acoustoelectric transport of electrons on helium (see Methods). The electron systems floats above the surface of a thin superfluid film at a temperature $T = 1.55$~K that is supported by an underlying piezoelectric substrate made of highly-polished lithium niobate (LiNbO$_{3}$). The superfluid film is estimated to be $\approx 70$ nm thick based on the amount of helium admitted into the experimental cell as well as its open volume (see Supplemental Information Section~1).  SAWs on the lithium niobate crystal are launched by applying a high-frequency voltage to an interdigitated transducer (IDT) on the surface of the lithium niobate and are directly detected using an opposing IDT (see Fig.~\ref{fig1}(a)). The electron system is trapped and laterally confined on the surface of the superfluid film using voltages applied to electrodes beneath and around the piezo-substrate. The underlying electrodes also serve to capacitively detect the signal produced when the evanescent electric field of the SAW carries electrons along the surface of the superfluid in its propagation direction. With this device we are able to perform both continuous wave (cw) and gated acoustoelectric measurements. For continuous wave acoustoelectric transport measurements an amplitude modulated (AM) excitation signal was applied to the exciter IDT and the acoustoelectric current $I_{\text{ae}}$ was measured using standard lock-in techniques. Time-of-flight measurements were performed by gating the SAW excitation signal on resonance (296 MHz) and the acoustoelectric current signal was collected using a fast, low-noise current preamplifier in tandem with a digital oscilloscope (see Methods).

The acoustoelectric response for the case of cw SAW excitation is shown in Fig.~\ref{fig1}(b) and coincides with the independently measured SAW resonance at 296 MHz (Fig.~\ref{fig1}(c)). This acoustoelectric current $I_{\text{ae}}$ is equivalent to the flux per unit time of electrons passing through the region above the underlying detection electrodes. No acoustoelectric current is observed without electrons present on the superfluid film (dashed blue trace in Fig.~\ref{fig1}(b)) confirming that the signal arises from the electron layer floating on the helium surface. Moreover, we find that $I_{\text{ae}}$ vanishes when the electrons are moved far away from the surface of the piezo-substrate by increasing the thickness of the superfluid layer (solid red trace in Fig.~\ref{fig1}(b)) as would be expected given the evanescent character of the SAW potential above the piezoelectric surface. The acoustoelectric current signal shown in Fig.~\ref{fig1}(b) also exhibits a periodic superimposed oscillation on top of the the main resonance. This corrugation in the acoustoelectric current is attributable to reflections of the SAW from the edge of the substrate, as is evident from the Fourier transform of the signal into the time domain (inset Fig.~\ref{fig1}(b)). We also note that the frequency of surface capillary wave excitations of the helium is expected to be $\sim 100$ kHz at the SAW wavelength, far from the $296$ MHz SAW resonance. More importantly, as described below our time-of-flight measurements clearly show that the acoustoelectrically transported electrons are propagating at the speed of the SAWs on lithium niobate ($\sim 10^{3}$~ m/s) and not of the much slower helium capillary waves ($\sim 10$~ m/s), excluding their contribution to the measured current. Together these experiments confirm that the acoustoelectric current is generated by the transport of electrons on the superfluid surface via the SAW electric field extending up from the underlying piezo-substrate.

The magnitude of this SAW-driven current depends on the extent to which the SAW potential can be screened by the electron system. In turn this will depend on the ratio of the plasma frequency, $\omega_{\text{p}}\approx15$~GHz in our experiments, to the SAW frequency. In this regime, where $\omega_{\text{p}} \gg \omega_{\text{SAW}}$ the mobile electrons on the surface of the helium will quickly respond to screen the piezo-field of the SAW, leading to a traveling charge density wave that produces the low-frequency acoustoelectric current we observe (see Supplemental Information Section 2). A characteristic feature of this acoustoelectric transport in 2DESs is a linear dependence of the measured signal, in this case $I_{\text{ae}}$, on the SAW intensity and hence excitation power (see Supplemental Information Section 2, Eq.~11). Fig.~\ref{fig2}(a) shows the SAW power dependence of $I_{\text{ae}}$ for electrons on helium. At the SAW resonance $I_{\text{ae}}$ increases linearly with the RF input power consistent with this expected linear response (see inset Fig.~\ref{fig2}(a)), which serves to further confirm the acoustoelectric origin of the measured signal from the system of electrons on helium.
\begin{figure}[hbtp]
\centerline{\includegraphics[height=0.3\textheight]{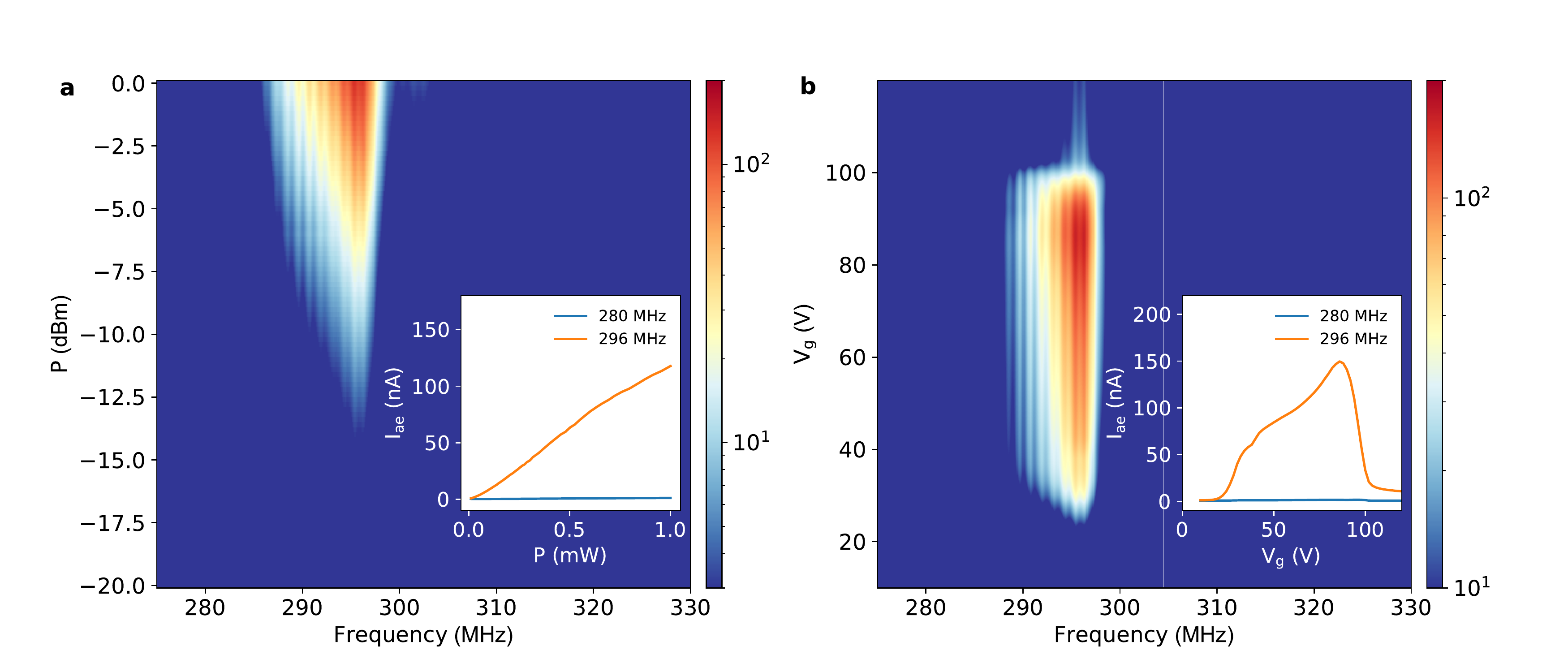}}
\caption{\label{fig2}{\bf{Power dependence and gate-tunability of the cw acoustoelectric effect with electrons on helium.}} \textbf{a)} Acoustoelectric current $I_{\text{ae}}$ measured as function of the SAW driving frequency and power. The inset shows that $I_{\text{ae}}$ is linear in driving power when the frequency corresponds to the SAW resonance. \textbf{b)} Demonstration of an acoustoelectric field effect transistor (aFET) with electrons on helium. The inset shows line-cuts of $I_{\text{ae}}$ both on- and off-resonance (orange and blue curves) with the SAW. These measurements were performed at $T=1.55$ K and with $V_{\text{s}}$ = $V_{\text{d}}$ = 40 V, corresponding to $n \cong 0.8 \times 10^9$~cm\textsuperscript{-2}, and the guard electrode biased with $-3.2$ V.}
\end{figure}

Field effect control is crucial for the development of acoustoelectric devices~\cite{Her11,McN11,Bar00,Pol16}. In Fig.~\ref{fig2}(b) we show a form of gate controlled SAW-driven electron transport. By tuning the gate bias voltage $V_{\text{g}}$ we can turn $I_{\text{ae}}$ ON and OFF, in effect creating an acoustoelectric field effect transistor (aFET). When $V_{\text{g}}$ is different from the source voltage $V_{\text{s}}$, electrons dragged by the traveling SAW encounter an effective potential energy barrier ${U_{\text{eff}}=-e(V_{\text{g}}-V_{\text{s}})}$ in the region above the gate. As shown in the orange trace of the Fig.~\ref{fig2}(b) inset, for sufficiently small values of $V_{\text{g}}$, acoustoelectric charge transport is blocked by a large positive $U_{\text{eff}}$, which results in zero current. Upon increasing $V_{\text{g}}$, electrons transported by the SAW are allowed to enter the region above the gate due to a decrease in $U_{\text{eff}}$, which leads to a increase in $I_{\text{ae}}$ at a threshold value of $V_{\text{g}}$ determined by the overall areal electron density. Further increasing $V_{\text{g}}$ eventually leads to a suppression of $I_{\text{ae}}$ once the region above the source has been depleted of electrons (see Fig.~\ref{fig2}(b) inset). 

\begin{figure}[hbtp]
\centerline{\includegraphics[height=0.5\textheight]{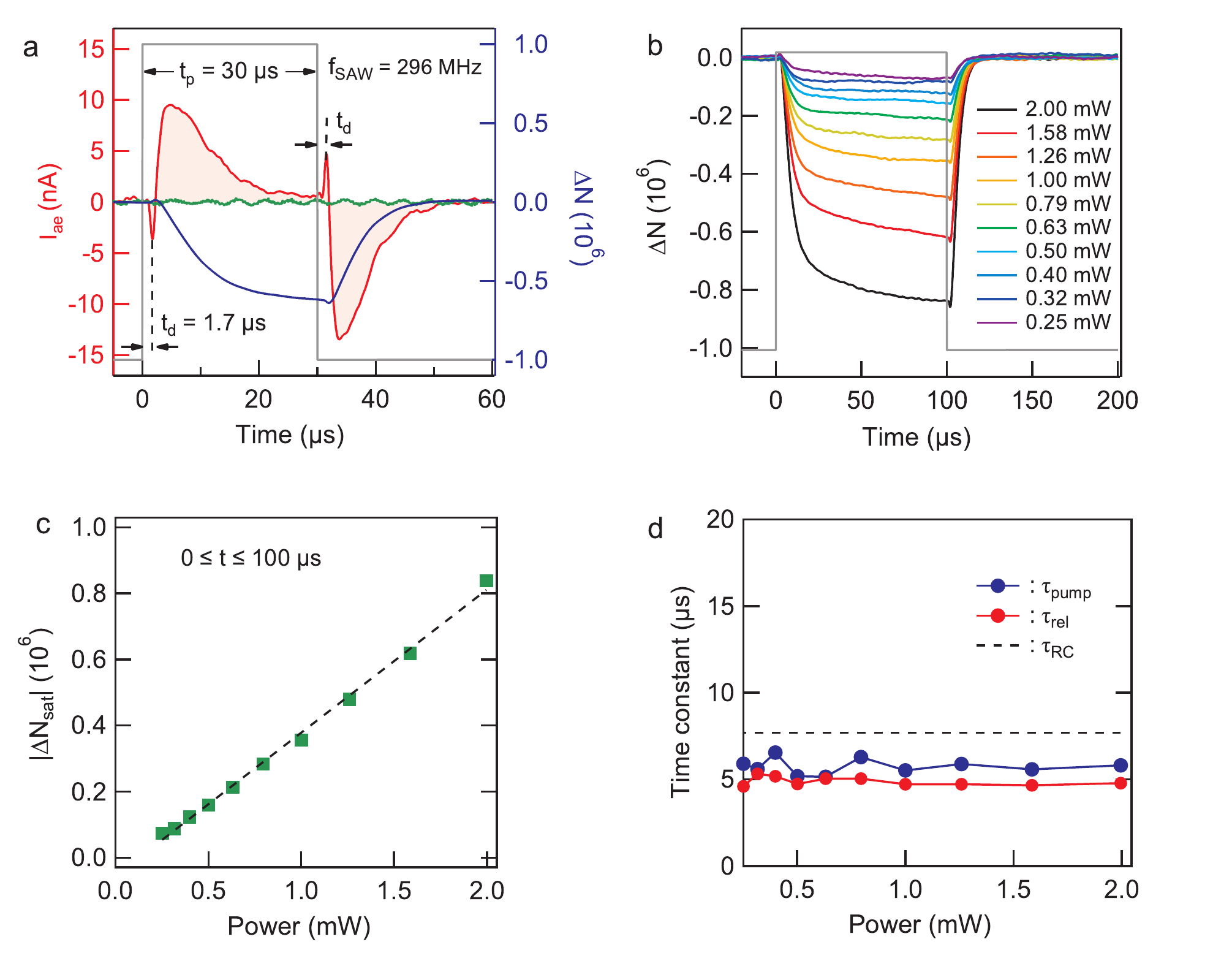}}
\caption{\label{fig3}{\bf{Time-of-flight acoustoelectric measurements of electrons on helium.}} \textbf{a)} Measurement of $I_{\text{ae}}$ (red curve) using a gated SAW (gray line) at a fixed RF power (0 dBm). These measurements were performed at the SAW resonance frequency (296 MHz). In these measurements the number of electrons above the detection electrode is $N\cong 0.4\times10^{9}$. The blue trace is the calculated change in the number of electrons $\Delta N$, which is obtained from the time integral of $I_{\text{ae}}$ (red trace). We note that $I_{\text{ae}}$ is absent when surface electrons are not present (green data trace), revealing that the measured current is induced by the SAW-electron interaction. \textbf{b}) $\Delta N$ and \textbf{c)} $|\Delta N_{\text{sat}}|$ for a steady-state SAW signal at various values of the driving RF power. The dashed line is a guide to the eye. In these measurements the number of electrons is $N \cong 0.8\times10^{9}$. \textbf{d)} Dynamical response of two-dimensional electrons on helium to gated SAWs as a function of RF power with $N\cong 0.8\times10^{9}$. The time constants $\tau_{\text{pump}}$ ($\tau_{\text{rel}}$) are those associated with the rising (falling) edge of the SAWs signal envelope and are determined from fitting $I_{\text{ae}}(t)$. The time constant $\tau_{\text{RC}}$ (dashed line) is that determined from low frequency transport of electrons on helium.}
\end{figure}

An important step for future SAW-based experiments with electrons on helium, in particular those working with small numbers of electrons, is the ability to precisely control the number of SAW-transported electrons. This can be accomplished by gating the SAW in time, which we show in Fig.~\ref{fig3}.
In these time-of-flight measurements the SAW IDT is excited on resonance for a fixed period of time, $t_{\text{p}}$, which launches a surface acoustic wave packet having a duration in time equal to $t_{\text{p}}$. The envelope of the SAW-packet contains the high-frequency acoustoelectric field, which picks up electrons and carries them in the propagation direction of the SAW (see Methods). For an areal electron density $n$ the continuity equation for the acoustoelectric current density~$\textbf{j}_{\text{ae}}$,
\begin{align}
\partial n / \partial t + \nabla \cdot \textbf{j}_{\text{ae}} = 0
\end{align}
predicts that an acoustoelectric current should appear once SAW-driven electrons flow past the boundary between the detecting electrodes. This behavior is shown in Fig.~\ref{fig3}(a) for the case where $t_{\text{p}} = 30~\mu$s, where $I_{\text{ae}}$ starts sharply increasing at a time $t_{\text{d}} \cong 1.7~\mu$s after the SAW is launched. This delay in the onset of $I_{\text{ae}}$ corresponds to the arrival time of the leading-edge of the gated SAW at the boundary above the detection electrode. The traveling SAW continues to drive electrons across the boundary, progressively building up an increasing charge imbalance in the electron layer. This charge imbalance produces an electric field that opposes the SAW-induced electron flow and leads to a decrease in $I_{\text{ae}}$ as shown in Fig.~\ref{fig3}(a) (red trace). Once the trailing edge of the gated SAW transits past the edge of the detection electrode (i.e. once $t = t_{\text{p}}$ + $t_{\text{d}}$) a current reappears but having the opposite polarity. In contrast to other 2DESs, these features are unique to electrons on helium and arise from the fact that this 2DES has a fixed total number of electrons. The SAW signal dynamically redistributes the electrons above the surface of the superfluid creating a non-equilibrium density, which then relaxes back to equilibrium after the passage of the SAW envelope.

We can quantitatively analyze this SAW-induced charge imbalance by calculating the time integral of $I_{\text{ae}}$ to extract the change in the number of electrons $|\Delta N|$ above the detection electrode (blue trace Fig.~\ref{fig3}(a)), which reaches a saturated steady state value, $|\Delta N_{\text{sat}}|$, in the limit of long SAW signal. Increasing the SAW power can be used to tune the magnitude of $|\Delta N_{\text{sat}}|$ as shown in Fig.~\ref{fig3}(b), and in this regime the power dependence allows us to estimate the minimum number of transported electrons that can be detected for a given length of the SAW envelope (see Fig.~\ref{fig3}(c)). For these measurements where $t_{\text{p}} = 100~\mu$s we find that for the lowest power as little as $|\Delta N/N|~\leq~10^{-4}$ of the number of electrons $N \cong 0.8 \times 10^9$ can be transported within each $100~\mu$s long SAW envelope (see Supplemental Information Section 3). We anticipate that optimization of device geometry to include a combination of microchannel lateral confinement~\cite{Rees2016} and single electron transistor charge detection~\cite{Schoelkopf1998,NRB2020} is a fruitful path forward for future experiments acoustoelectric experiments with even smaller numbers of electrons.

Finally, the gated SAW measurements allow us to extract information about the dynamical response of the many-electron system on helium by analyzing the build-up and relaxation of the SAW-induced charge pumping. Specifically, we fit the rising and falling edges of the acoustoelectric signal to an exponential and extract time constants $\tau_{\text{pump}}$ and $\tau_{\text{rel}}$, which are shown in Fig.~\ref{fig3}(d). In both cases, the time constants do not depend on the SAW power and we find $\tau_{\text{pump}} = 5.7 \pm 0.1$ $\mu$s and $\tau_{\text{rel}} = 4.9 \pm 0.1$ $\mu$s (see Supplemental Information Section 4). These time constants are ultimately determined by the underlying mechanisms that lead to electron scattering in the system. In the case of the present experiments, which were performed at $T=1.55$ K, a strong scattering mechanism is the collision of electrons with helium vapor atoms above the superfluid surface. These scattering events occur at a frequency ($\approx$~18~GHz, see Supplemental Information Section 3) much larger than that of the SAW electric field (296 MHz). Therefore the time constants extracted from the SAW measurements should be similar in magnitude to the $RC$ time constant expected from a transmission line modeling of the low-frequency transport measurements of the electron system, which we performed using the electrodes located underneath the LiNbO$_{3}$ substrate (see Supplemental Information Section 3). Based on a transmission line analysis of these transport measurements we estimate that $\tau_{\text{RC}} = 7.7$~$\mu$s, in reasonable agreement with our SAW measurements.

In conclusion, we have demonstrated the coupling of a system of electrons floating on the surface of superfluid helium to the evanescent field produced by a piezoelectric surface acoustic wave. This method allows us to perform controlled high-frequency acoustoelectric charge pumping within the electron system. These measurements also show that SAWs are a versatile tool for interrogating the dynamical processes in electrons on helium including, when extended to lower temperature, investigations of plasmon modes and ripplo-polaronic excitations of the Wigner solid~\cite{Badrutdinov2020}.

\section{Methods}
The electrons on helium acoustoelectric device was measured inside a superfluid leak-tight copper cell, where a sheet of electrons produced from a tungsten filament were floating on top of $\sim$ 70 nm thick liquid helium film. A YZ cut lithium niobate (LiNbO\textsubscript{3}) single crystalline chip with a length of 20 mm, a width of 10 mm, and a thickness of 0.5 mm was used as a substrate for both the superfluid film and SAW propagation as illustrated in Fig.~\ref{fig1}(a). A set of three rectangular electrodes (source, gate, and drain) located beneath the lithium niobate as well as a rectangular guard electrode outside of the substrate were used to trap and laterally confine electrons above the surface of the helium film by applying DC bias voltages to them. Each of the underlying trapping electrodes had a width of 4.95 mm and a length of 9 mm. On the surface of the LiNbO\textsubscript{3} chip, two identical inter-digitated transducers (IDT) consisting of 40 pairs of 3 $\mu$m wide fingers were patterned using standard optical lithography. The transducers were made of evaporated aluminum and had a thickness of approximately 70 nm and a width of 4 mm, which corresponds to the aperture of the SAW beam. By applying a high frequency signal to the exciter IDT, SAWs are launched along the surface of the piezoelectric substrate toward the opposing detector IDT. The fundamental resonant frequency of our SAW device is calculated to be $\nu = \lambda/v$ = 291 MHz, dictated by the IDT finger periodicity, $\lambda$ = 12 $\mu$m, and the speed of sound in YZ cut LiNbO\textsubscript{3}, $v$ = 3488 m/s. We characterized the frequency response of the SAW used in our experiments using a vector network analyzer. Fig.~\ref{fig1}(d) shows the measured transmission coefficient $S_{12}$ of the SAW delay line as function of frequency at ${T = 1.55}$ K. The resonance in the transmitted power at $\nu$~=~296 MHz is associated with the generation of SAW on the substrate. The slight difference between the expected resonant frequency and the measured value is likely due to piezo-crystal contraction at cryogenic temperature. With this experimental setup, acoustoelectric transport of electrons on helium was measured via capacitive coupling between electrons floating on the helium surface and a detection electrode beneath the lithium niobate. For continuous wave (cw) acoustoelectric transport measurements an amplitude modulated cw excitation signal having a modulation frequency of 20 kHz was applied to the exciter IDT using an Agilent 8648B RF signal generator and $I_{\text{ae}}$ was measured using standard lock-in techniques. The time-of-flight measurements of $I_{\text{ae}}$ were performed by gating the SAW excitation signal on resonance (296 MHz) and the acoustoelectric current signal was collected using an SR570 low noise current preamplifier in tandem with a Tektronix DPO7054 digital oscilloscope. The acoustoelectric current waveform was averaged over 10,000 samples to improve signal to noise ratio in these measurements.

\section{Data Availability}
The data that support the findings of this study are available from the corresponding authors upon reasonable request.


\section{Acknowledgments}
We are grateful to M.I. Dykman, D.I. Schuster, D.G. Rees, K. Kono, J. Kitzman, and C.~Mikolas for illuminating and fruitful discussions. We also thank B. Bi for technical assistance and use of the W.M. Keck Microfabrication Facility at MSU. This work was supported by the National Science Foundation via grant numbers DMR-1708331 and DMR-2003815. J.P., J.R.L. and L.Z. acknowledge the valuable support of the Cowen Family Endowment at MSU and N.R.B. acknowledges the support of a sponsored research grant from EeroQ Corp.

\section{Author Contributions}
H.B. fabricated the SAW devices with assistance from R.L. and also built the SAW-based measurement setup. H.B. and K.N. performed the experiments with assistance from J.R.L., N.R.B. and L.Z. H.B. analyzed the data. J.P. conceived of the experiments, supervised the project and provided guidance. All authors contributed to the writing the manuscript.

\section{Competing Interests}
J. Pollanen is a co-founder and partial owner of EeroQ Corp. All other authors declare no competing interests.

\end{document}


\makeatletter 
\renewcommand{\thefigure}{S\@arabic\c@figure}
\makeatother

\section*{Supplemental Information}
\subsection*{1. Superfluid film thickness}
\begin{figure}[hbtp]
  \includegraphics[width=0.8\textwidth]{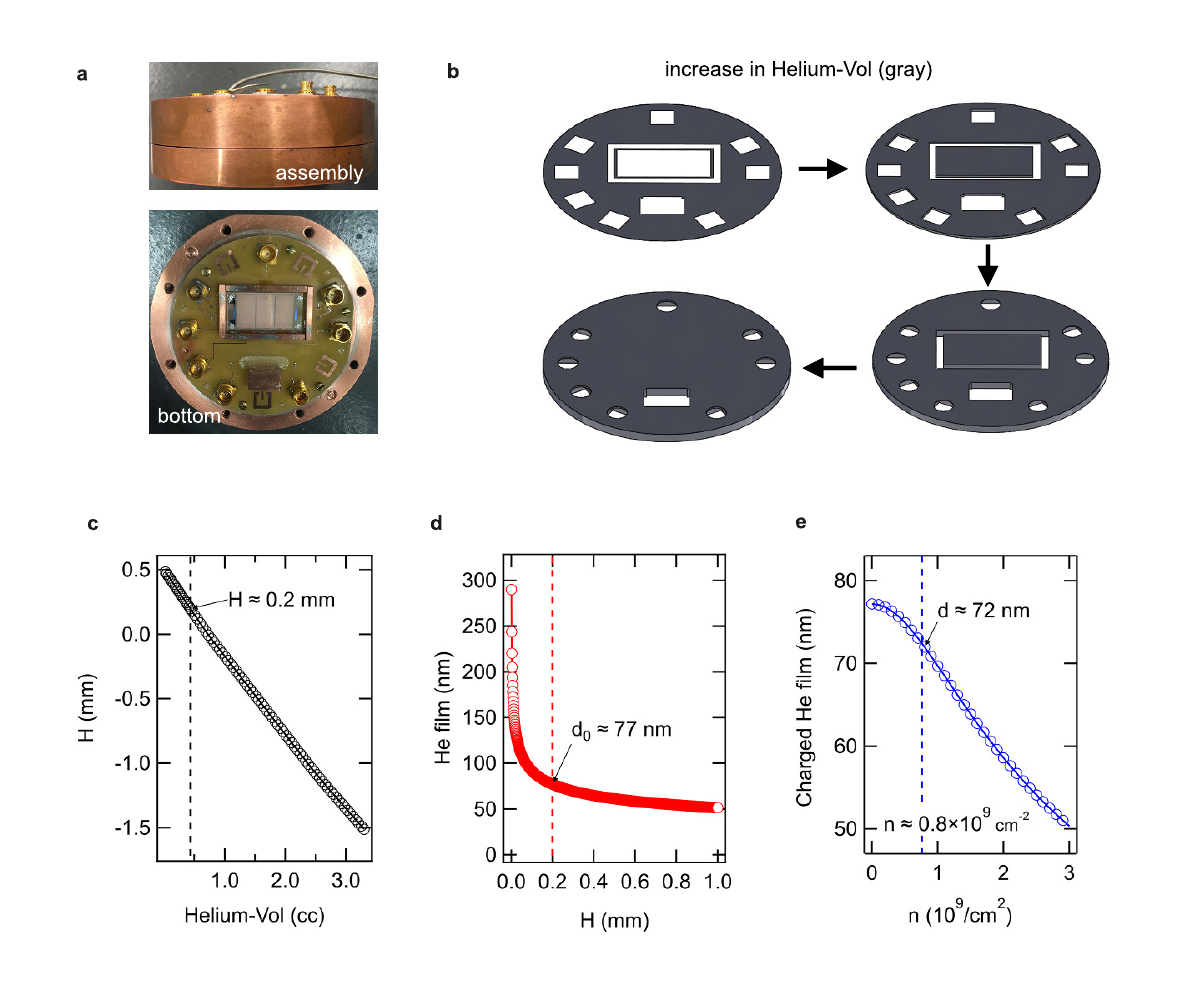}
\caption{\textbf{The experimental sample cell and helium film thickness determination.} \textbf{(a)}~Photograph of the copper cell used for the SAW-based measurements of electrons on helium. The bottom portion of the cell contains a LiNbO\textsubscript{3} device (semi-transparent white) mounted on a printed circuit board \textbf{(b)} 3D CAD modeling of the liquid helium volume inside the experimental cell during filling with helium. \textbf{(c)} Calculation of $H$, the distance between the piezo-substrate top and the liquid helium level in the cell reservoir volume from the 3D CAD modeling of the open cell volume. The calculated value of $H$ in the experiments is 0.2 mm for a helium vol.~=~0.44~cc (vertical dashed black line). \textbf{(d)} Helium film thickness versus $H$ before charging the film (see~Eq.~\eqref{neutral}). For an uncharged helium film $d_{0}$ $\approx$ 77 nm (vertical dashed red line). \textbf{(e)} Charged helium film thickness versus electron density from Eq.~\eqref{charged}. The charged helium film thickness is calculated to be $d$ $\cong$ 72 nm for an electron density of $n \cong 0.8 \times 10^{9}$ cm\textsuperscript{-2} (vertical dashed blue line).}
\label{figS1}
\end{figure}

Fig.~\ref{figS1}(a) shows the hermetically sealed copper cell used for the measurements described in the main manuscript. To fill the cell with superfluid \textsuperscript{4}He, helium gas was supplied into the cell at $T \cong$ 1.55 K through a capillary line at room temperature. The liquid helium volume admitted into the cell was determined by varying the pressure in a calibrated standard volume ($V = 260$ cc) located at room temperature. The thickness of a saturated helium film, $d_{0}$, can be estimated from~\cite{pobell2007matter},
\begin{equation}
\label{neutral}
 \frac{\alpha}{{d_{0}}^4} = {\rho g H},
\end{equation}
where $\alpha$ is the van der Waals constant, $g$ is the gravitational acceleration, $\rho$ is the mass density of helium, and $H$ is the distance from the LiNbO\textsubscript{3} substrate top surface down to the liquid helium surface in the reservoir volume inside the cell. We calculated $H$ as function of the volume of helium admitted into the cell using 3D modeling of the cell open volume (see~Fig.~\ref{figS1}(b),(c)). For the thin film measurements reported in the manuscript, $H~=~0.2$~mm, which corresponds to a thickness of the superfluid film, $d_{0} \cong$ 77 nm, from Eq.\eqref{neutral}. Charging the helium film with electrons exerts an electronic pressure on the film~\cite{leiderer1992electrons} in addition to gravity and thus reduces the thickness of the charged helium film, $d$, which can be calculated from,
\begin{equation}
\label{charged}
  \frac{\alpha}{{d}^4} = {\rho g H} + p_{\text{el}} = \frac{\alpha}{{d_{0}}^4} + 2\pi {n}^2 {e}^2,
\end{equation}
where ${p}_{\text{el}}$ is the electronic pressure, $n$ is the electron density and $e$ is the electron charge. We note that due to quantum electrodynamic effects, for thick films ($>$ 60 nm) the van der Waals potential energy is proportional to $1/{d^4}$ rather than $1/d^3$~\cite{takita2015electrons,klier2003equilibrium}. In most of the acoustoelectric measurements reported in the manuscript, the areal electron density is estimated to be $n \simeq 0.8 \times 10^{9}$ cm\textsuperscript{-2} from field effect transistor (FET) measurements, which are described in the SI section 3. This yield a charged helium film thickness of $d~\cong$~72~nm from Eq.~\eqref{charged}.

\clearpage
\subsection{2. SAW-2DES interaction and acoustoelectric transport}
As a SAW propagates along the surface of a piezoelectric substrate, co-propagating electric fields can interact with a 2D electron system in close vicinity to the substrate surface. The electrons dynamically reorder themselves to accommodate the spatially and temporally varying SAW piezo-field. Via reciprocity, this interaction influences the propagation of SAW and leads to attenuation per unit length $\Gamma_{\text{el}}$ and a shift of the SAW velocity $\Delta v/v_{0}$ given by~\cite{Wix89}
\vspace{1mm}
\begin{equation}
\label{attenuation}
 \Gamma_{\text{el}} = k\frac{K^{2}}{2}\frac{\sigma/\sigma_{\text{m}}}{1+(\sigma/\sigma_{\text{m}})^{2}}
\end{equation}\vspace{-0.5mm}
\begin{equation}
\label{velocity}
\frac{\Delta v}{v_{0}}=\frac{v-v_{0}}{v_{0}}=\frac{K^{2}}{2}\frac{1}{1+(\sigma/\sigma_{\text{m}})^{2}},
\end{equation}\vspace{-3mm}

\noindent where $k$ is the wave number of the SAW, $\sigma$ is the sheet conductivity of the 2DES, $K^{2}$ is an effective electromechanical coupling coefficient, $\sigma_{\text{m}}$ is a characteristic conductivity that depends on material parameters, and $v_{0}$ is the SAW velocity on the free surface of the piezo-substrate in the absence of a 2DES.

Additionally, the ability of the SAW potential to trap and transfer electrons at the speed of sound in the substrate enables the generation of an acoustoelectric current ($I_{\text{ae}}$) through the 2DES. The piezoelectric field induced by a SAW propagating along the $x$-axis is given by $E_{\text{p}} (x,t) = Ae^{i(kx - \omega t)}$, where $\omega \text{ and } k$ are the frequency and wavevector of the SAW and $A$ is the amplitude fo the field. The dynamical response of the electron system acts to screen this potential to an extent dictated by the ratio of the plasma frequency to the SAW frequency. For an electron density is $n \sim 10^{9}$ cm$^{-2}$ and the SAW wavevector $k \cong 0.5 \times 10^{6}$~m$^{-1}$ the 2D plasma frequency is $(e^{2} n / 2 \varepsilon_{0} m)^{1/2} k^{1/2} / 2 \pi \cong 15$ GHz, which is much larger than the SAW frequency $\omega_{\text{SAW}} / 2 \pi \cong$ 0.3 GHz. Thus, in our experiments the electrons on the surface of liquid helium effectively screen the piezoelectric field. In this regime the effective field becomes~\cite{Wix89,ESSLINGER1994}
\begin{equation}
E_{\text{eff}} (x,t) = E_{\text{p}} (x,t) + E_{\text{ind}} (x,t) = \dfrac{E_{\text{p}} (x,t)}{1 + i (\sigma / \sigma_{\text{m}})}
\end{equation}
and is accompanied by a modulated surface electron density
\begin{equation}
n(x,t) = n_{0} + \Delta n \, e^{i (kx - \omega t)} = n_{0} + \Delta n (x,t),
\end{equation}
where $E_{\text{ind}}(x,t)$ is the screening field produced by the response of the electron system and $n_{0}$ is the equilibrium electron density. In the case where the SAW produces relatively small disturbances in the electron density (valid in this experiment), the modulated conductivity of the electron system may be expanded to first order
\begin{equation}
\sigma(x,t) = \sigma_{0} + \dfrac{\partial \sigma}{ \partial n} \, \Delta n (x,t)
\end{equation}
with the local acoustoelectric current density given by
\begin{equation}
j_{\text{ae}} (x,t) = \sigma E_{\text{eff}} (x,t) = \sigma_{0} E_{\text{eff}} (x,t) + \dfrac{\partial \sigma}{\partial n}~\Delta n (x,t)~E_{\text{eff}} (x,t).
\label{eq:jae}
\end{equation}
From the continuity equation, $\partial_{x} \, j_{\text{ae}} + \partial_{t} \, (-ne )  = 0$, one can relate the oscillating charge density $\Delta n (x,t)$ to the effective field $E_{\text{eff}} (x,t)$, i.e.
\begin{equation}
 \Delta n (x,t) = - \dfrac{\sigma E_{\text{eff}} (x,t)}{ev},
\label{eq:deltan}
\end{equation}
where $e$ and $v = \omega / k$ are electron charge and the SAW velocity. By substituting Eq.~\eqref{eq:deltan} into Eq.~\eqref{eq:jae} $j_{\text{ae}}$ can be seen to have both linear and quadratic dependence on the effective potential $E_{\text{eff}} (x,t)$
\begin{equation}
j_{\text{ae}} (x,t) = \sigma_{0} E_{\text{eff}} (x,t) - \dfrac{\partial \sigma}{\partial n} \dfrac{\sigma_{0}}{e v} E_{\text{eff}}^{2} (x,t).
\label{eq:jae2}
\end{equation}
Our low-frequency (quasi-DC) measurements of the acoustoelectric current arise from the time average of this current density (i.e. $\langle j_{\text{ae}} (x,t) \rangle_{t} $), in which only the quadratic term $E_{\text{eff}}^{2} (x,t)$ contributes to the signal since $\langle E_{\text{eff}} (x,t) \rangle_{t} = 0$. This results in the acoustoelectric current depending linearly the SAW intensity $I(x) = \sigma_{0} \langle E_{\text{eff}}^{2} (x,t) \rangle_{t} / \Gamma_{\text{el}}$, i.e.
\begin{equation}
\langle j_{\text{ae}}(x,t) \rangle_{t} = \dfrac{\mu}{v} \Gamma_{\text{el}} I(x),
\end{equation}
where $\mu$ is electron mobility.

The intensity of the SAW can be expressed as $I(x) = (\alpha_{\text{L}} P_{\text{in}} e^{-\Gamma_{\text{tot}} x}) / w$, where $\alpha_{\text{L}}$, $P_{\text{in}}$, $w$, and $\Gamma_{\text{tot}}$ denote the acoustoelectric conversion efficiency of the SAW-exciting IDT, the input RF power applied to the IDT, the SAW beam aperture, and the total attenuation per unit length of SAWs respectively. We note that the device used here is composed of 2D electrons floating above $\approx 70$ nm thick helium film formed on a LiNbO\textsubscript{3} substrate. Such a helium film on LiNbO\textsubscript{3} is known to lead to additional SAW attenuation $\Gamma\textsubscript{He}$~\cite{bye19}. Thus, the total attenuation of the SAW for the electrons on helium system is given by
\begin{equation}
 \label{tot attenuation}
\Gamma_{\text{tot}} = \Gamma_{\text{el}} + \Gamma_{\text{He}}.
\end{equation}

\clearpage
\subsection*{3. Characterization of the electron system on helium}
\subsubsection*{Field effect transistor transport measurements}
\begin{figure}[hbtp]
  \includegraphics[width=0.8\textwidth]{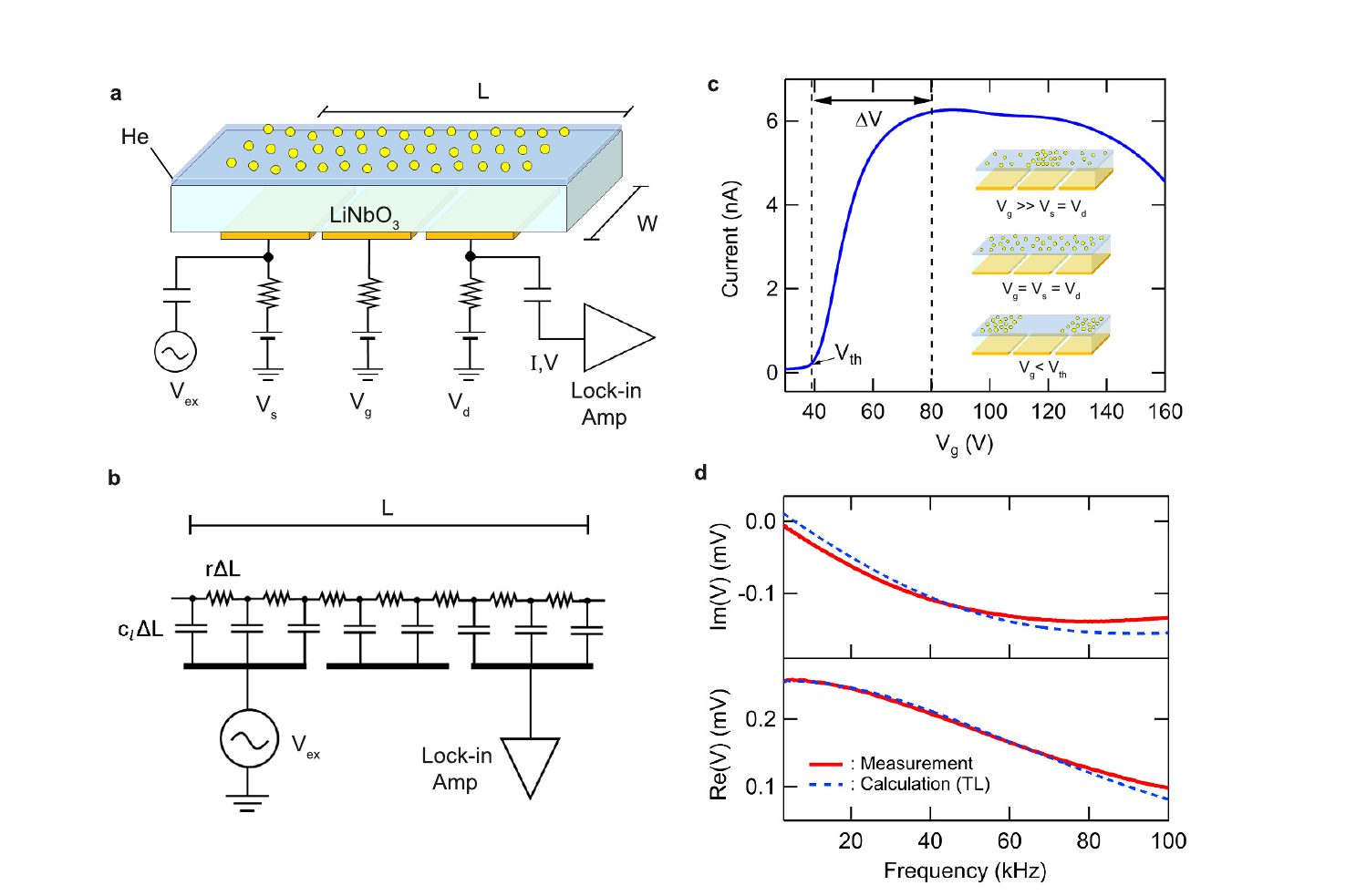}
\caption{\textbf{Experimental setup, circuit model, and measurement data for low-frequency transport of electrons on helium.} \textbf{(a)} Sketch of the electrons on helium device. A DC bias voltage is applied to the three underlying electrodes to trap electrons above helium surface. An AC excitation voltage ($V_{\text{ex}}$ = 0.1 V) applied to the source electrode induces a current through electrons on helium, which is capacitively detected using the drain electrode. \textbf{(b)} The equivalent transmission line circuit model for this device. The resistance of the electron sheet and the capacitance between this sheet and the three electrodes are spatially distributed per unit length, $r \Delta L$ and $c_{l} \Delta L$, where $r$ and $c_{l}$ represent the resistance per unit length and the capacitance per unit length. \textbf{(c)} Current-voltage characteristics for FET operation of the electrons on helium device. For this representative data, the source-drain current amplitude was measured during a gate voltage sweep with $V_{\text{s}}$~=~$V_{\text{d}}$~=~80V. The insets show the corresponding electron density profiles over the electrodes for different values of $V_{\text{g}}$. A uniform areal density distribution is achieved at $V_{\text{g}}$~=~$V_{\text{s}}$~=~$V_{\text{d}}$. \textbf{(d)} Frequency dependence of the transport characteristics of the system of electrons on helium. The red and blue traces represent the experimental data and the calculation based on the transmission line model respectively. For these measurement, all three electrodes were biased with 80 V ($V_{\text{g}}$~=~$V_{\text{s}}$~=~$V_{\text{d}}$~=~80 V).}
\label{figS2}  
\end{figure}

Fig.~\ref{figS2}(a) and (b) illustrate the measurement set-up and the equivalent transmission line circuit model for the low-frequency transport properties of electrons on helium. After charging a helium film with electrons, we characterize the electron distribution with low-frequency field effect transistor (FET) operation of the device~\cite{nasyedkin2018unconventional}. These measurements are done with a gate voltage $V_{\text{g}}$ sweep at fixed source and drain voltage at a frequency of 60 kHz as shown in Fig.~\ref{figS2}(c). For sufficiently small values of the gate voltage, $V_{\text{g}}$ $<$ $V_{\text{th}}$, all of the electrons are localized over the source and drain electrodes creating a depletion region above the gate, which leads to zero current through electron layer. At a threshold value of the gate $V_{\text{g}} = V_{\text{th}}$ an AC current onsets from electrons being attracted to the region above the gate. As $V_{\text{g}}$ increases beyond $V_{\text{th}}$, source-drain current quickly rises and reaches a maximum in a vicinity of uniform electron density ($V_{\text{s}} = V_{\text{d}} = V_{\text{g}}$). After reaching its maximum value, the current begins decreasing and eventually vanishes as electron depletion over the source and drain electrode occurs. This ability to change the electron density above the individual electrodes with the application of a bias voltage enables the electrically switching ON and OFF of the acoustoelectric current in electrons on helium system (see Fig.~2(b) of the main manuscript).

 \vspace{-5mm}

\subsubsection{Electron density}
A homogeneous areal electron density $n$ above all electrodes is achieved when $V_{\text{s}}$~=~$V_{\text{g}}$~=~$V_{\text{d}}$ and the electron density can be calculated from the low-frequency FET transport data by measuring the difference $\Delta V$ between the case where the electron density is uniform and $V_{\text{th}}$ (see~Fig.~\ref{figS2}(c))~\cite{nasyedkin2018unconventional},
\begin{equation}
\label{n}
  n = \frac{2~c_{l}}{3~e}\Delta V.
\end{equation}
Here, $c_{l}$ is the capacitance per unit area between electron layer and underlying electrodes from geometrical configuration, which is given by
\begin{equation}
 \label{cap}
  c_{l} = {\epsilon}_{0}\left(\frac{1}{\frac{d_{\text{g}}}{\epsilon}+\frac{d_{\text{s}}}{\epsilon_{\text{s}}}+\frac{d}{\epsilon}}\right) \cong {\epsilon}_{0}\left(\frac{1}{\frac{d_{\text{g}}}{\epsilon}+\frac{d_{\text{s}}}{\epsilon_{\text{s}}}}\right) \hspace{5mm} \text{(for $d \ll d_{\text{g}}$),}
\end{equation}
where $\epsilon_{0}$ is the vacuum permittivity, $\epsilon = 1.057$ is the dielectric constant of helium, $\epsilon_{\text{s}} \cong 35$ is the effective dielectric constant of the LiNbO\textsubscript{3},  $d_{\text{g}}$ $\cong$ 70 $\mu$m is the  gap between the LiNbO\textsubscript{3} substrate and the bottom electrodes, $d_{\text{s}}$ = 0.5 mm is the thickness of the LiNbO\textsubscript{3} substrate, and $d$ is the charged helium film thickness. The effective dielectric constant $\epsilon_{\text{s}}$ for LiNbO\textsubscript{3} is given by $\epsilon_{\text{s}}=\sqrt{\epsilon_{11}\epsilon_{33}-\epsilon_{13}^2}$ where $\epsilon_{11}$, $\epsilon_{33}$, and $\epsilon_{13}$ are the dielectric tensor element of LiNbO\textsubscript{3} at constant stress~\cite{muller2005surface}. These dielectric elements are given in Ref.~\cite{jazbinvsek2002material} as $\epsilon_{11}=44.3$, $\epsilon_{33}=27.6$, and $\epsilon_{13}=0$. With $c_{l} \cong 1.08 \times 10^{-7}$ F/m\textsuperscript{2} obtained from Eq.~\eqref{cap}, the uniform areal electron density is found to be $n \simeq 0.8 \times 10^{9}$ cm\textsuperscript{-2} for most of acoustoelectric measurements reported in this manuscript. For the power dependent pulsed SAW measurements (Fig.~3(b)-(d)), the value of $n$ is about 1.9 $\times 10^{9}$ cm\textsuperscript{-2}. We note that in the gate-tunable acoustoelectric measurements shown in Fig.~2(b), $n$ is no longer a uniform density but rather varies as a function of $V_{\text{g}}$, however the total number of electrons over the device is fixed. \vspace{0mm}

\subsubsection{Electron mobility}
Using the electrodes located beneath the LiNbO$_{3}$ substrate we are able to perform low-frequency transport measurements of the system of electrons on the helium thin film. The conductivity $\sigma$ of electrons on helium from our low-frequency transport measurements can be determined by fitting the frequency response measurement to a RC transmission line model~\cite{lea1991ac,mehrotra1987analysis} with $\sigma$ as fitting parameter~\cite{nasyedkin2018unconventional}. Fig.~\ref{figS2}(d) shows this frequency response for the representative case where $V_{\text{s}}$ = $V_{\text{g}}$ = $V_{\text{d}}$ = 80 V, which shows good agreement with this model fit. The value of the conductivity from this fit is $\sigma$ $\cong$ $1.58 \times 10^{-6}$ $\Omega^{-1}$. The mobility $\mu$ of electrons on helium is then calculated based on a Drude model analysis ($\sigma = ne\mu$). For an electron density of $n = 1.9 \times 10^{9}$ cm\textsuperscript{-2} the mobility is $\mu$ $\cong$ $5.3 \times 10^{3}$ cm$^{2}$/Vs with $d \cong 60$ nm. This mobility is roughly two orders of magnitude less than the mobility of electrons on bulk \textsuperscript{4}He at the same temperature~\cite{Iye80}. Such a low mobility for electrons on a thin superfluid film is consistent with previous measurements and can be explained by the close proximity of the electron layer to the underlying substrate~\cite{shikin2001dip}. 

\newpage
\subsubsection{RC delay time constant}
The RC delay time constant, $\tau_{\text{RC}}$, introduced in the manuscript is calculated based on the transmission line modeling described above (see Fig.~\ref{figS2}(b)) and the Elmore delay model~\cite{rabaey2002digital}, which is encapsulated by the following equation
\begin{equation}
\label{tau}
  \tau_{\text{RC}} = rc_{l}L^{2}\frac{1+N}{2N} \cong \frac{R_{\text{tot}}C_{\text{tot}}}{2} \hspace{5mm} \text{(for large N)},
\end{equation}
where $r$ and $c_{l}$ are the resistance per unit length and the capacitance per unit length of the electron system and $L$, $R_{\text{tot}}$, $C_{\text{tot}}$, and $N$ represent the total length, the total resistance, the total capacitance, and the number of the node in the entire transmission line. $R_{\text{tot}}~=~\sigma^{-1}W^{-1}L$ and $C_{\text{tot}}$~=~$c$~$WL$ are approximately $1.06$ M$\Omega$ and $14.6$ pF for $W~=~9$~mm (the width of the electrodes in the SAW device) and $L = 15$ mm, which yields $\tau_{\text{RC}} \approx 7.7$ $\mu$s in the low-frequency transport regime.

\subsubsection{Electron collision rate with helium vapor atoms}
At a temperature $T =1.55$ K, \textsuperscript{4}He vapor atoms are a strong source of electron scattering. The collision rate $1/\tau_{\text{He}}$ is determined by 
\begin{equation}
\dfrac{1}{\tau_{\text{He}}} = \dfrac{e}{\mu m^{*}}
\end{equation}
with the effective electron mass $m^{*} = m_{\text{e}}$ (bare electron mass) and the electron mobility $\mu \cong 1 \times 10^{5}$ cm$^{2}$/Vs~\cite{Iye80}. This yields $1/\tau_{\text{He}} \simeq 18$ GHz.

\subsection{4. Determination of SAW charge pumping time constants}
\begin{figure}[hbtp]
\centering
  \includegraphics[width=0.9\textwidth]{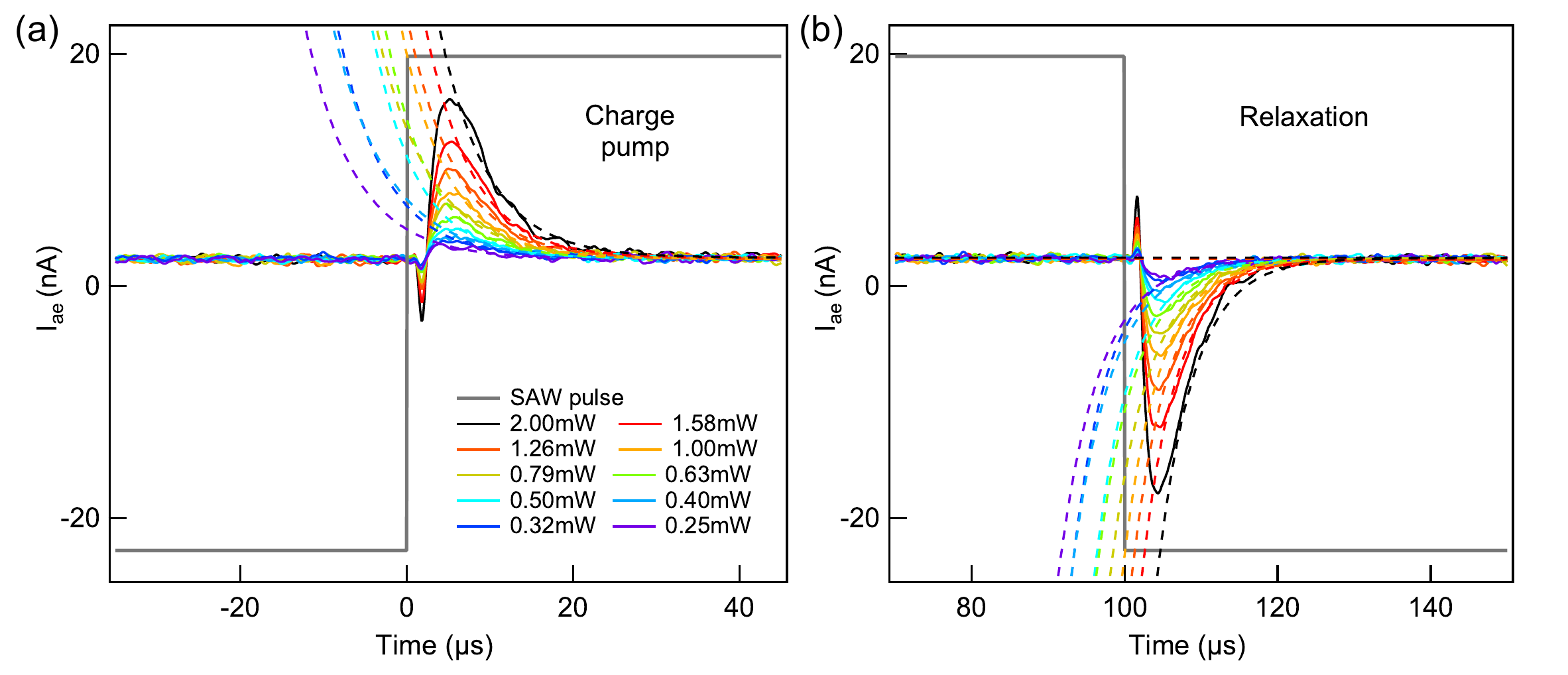}
\caption{\textbf{Determination of acoustoelectric time constants}. These time constants were determined from exponential fits to the acoustoelectric current data during \textbf{(a)} SAW excitation (charge pumping) and \textbf{(b)} relaxation after the SAW drive was removed.}
\label{figS3}
\end{figure}

The time constants for SAW-driven charge pumping ($\tau_{\text{pump}}$) and subsequent relaxation ($\tau_{\text{rel}}$) for different RF power was determined from an exponential curve fitting of the acoustoelectric current ($I_{\text{ae}}$) data as shown in Fig.~\ref{figS3} and resulting values are tabulated in the table below.

\begin{table} \vspace{-65mm}
 \centering 
 \resizebox{0.3\linewidth}{!}{
 \begin{tabular}{ccc} 
 \\[0.1ex]
 \hline\hline 
 RF (mW) & $\tau_{\text{pump}}$ ($\mu$s) & $\tau_{\text{rel}}$ ($\mu$s)\\
\hline 
0.25 & 5.9 & 4.6\\ 
0.32 & 5.6 & 5.3\\
0.40 & 6.5 & 5.2\\ 
0.50 & 5.2 & 4.7\\
0.63 & 5.2 & 5.0\\
0.79 & 6.3 & 5.0\\
1.00 & 5.5 & 4.7\\
1.26 & 5.9 & 4.7\\
1.58 & 5.6 & 4.7\\
2.00 & 5.8 & 4.8\\
\hline\hline 
\end{tabular}}
\caption{\textbf{Acoustoelectric time constants}. Values of time constants obtained from the exponential curve fits in Fig.~\ref{figS3} with varying SAW excitation power.}
\label{tab:tau}
\end{table}